\documentclass[12pt,a4paper]{article}

\usepackage{amsmath, amssymb, amsfonts}
\usepackage{bm}
\usepackage{geometry}
\usepackage{cite}
\usepackage{hyperref}
\hypersetup{colorlinks=true, linkcolor=blue, citecolor=blue, urlcolor=blue}
\usepackage{graphicx}
\usepackage{appendix}

\newcommand{\rr}{\mathbf{r}}
\newcommand{\kk}{\mathbf{k}}
\DeclareMathOperator{\diag}{diag}

\title{\textbf{A Spectral-Domain Pseudo-Inverse Method for True 3D Gravity Inversion}}
\author{CHEN Shengchang}
\date{School of Earth Sciences, Zhejiang University, Hangzhou, China \\ \texttt{chenshengc@zju.edu.cn}}

\begin{document}

\maketitle

\begin{abstract}
The main difficulty in 3D gravity inversion is that surface observations lack vertical wavenumber information, making the problem underdetermined and depth resolution poor. Building on the author's spectral-domain pseudo-inverse theory for unitary diagonalizable systems, this paper presents a true 3D inversion method. Surface data are analytically continued upward via Laplace's equation to form a 3D data volume, providing the vertical wavenumber sampling for the 3D Fourier transform. A general analytical expression for the half-space spectrum is derived, separating the horizontal spectrum from the vertical propagation kernel. The Green's function of the 3D Poisson equation is unitarily diagonalized, yielding the forward spectral response $\lambda(\kk) = -i 4\pi G k_z / K^2$, and a stable inverse filter $q_\alpha(\lambda) = \bar{\lambda}/(|\lambda|^2 + \alpha)$ is constructed. This filter is proved to have bounded stability and consistency (reducing to the exact inverse as $\alpha \to 0^+$). Validation with a homogeneous sphere model shows correct recovery of the anomaly location and singular behavior at the source. The contributions are twofold: (1) upward continuation constructs a 3D volume from 2D surface data, which does not add physical observations but unfolds the 3D spatial variation information (data innovation) derived through the governing physical equation; (2) the inversion is reduced to one forward 3D Fourier transform, one spectral scaling, and one inverse transform. The framework extends to any linear potential-field inverse problem with a translation-invariant forward operator, including magnetic inversion.

\end{abstract}

\section{Introduction}

Gravity exploration aims to infer subsurface density distributions from surface gravity measurements. Its mathematical foundation is the linear inverse problem associated with the 3D Poisson equation. Let the subsurface density distribution be $\rho(\rr)$ and the gravitational potential be $U(\rr)$. They satisfy
\begin{equation}
\nabla^2 U(\rr) = -4\pi G \rho(\rr), \tag{1}
\end{equation}
where $G = 6.674 \times 10^{-11} \, \text{m}^3\text{kg}^{-1}\text{s}^{-2}$ is the gravitational constant.

The observed vertical gravity anomaly is $g_z = -\partial U/\partial z$. The inverse problem is: given $g_z(x,y,0)$ on the surface, recover $\rho(x,y,z)$ in the subsurface half-space.

This inverse problem faces three major challenges. First, \textbf{underdetermination}: the data are available only on a 2D plane, while the density distribution is 3D, leading to an insufficient number of independent constraints. Second, \textbf{instability}: the gravity field decays rapidly with depth, so signals from deeper sources are severely attenuated at the surface, making the inversion highly ill-conditioned and sensitive to noise. Third, \textbf{poor depth resolution}: data from a single depth plane carry no vertical wavenumber information, resulting in limited resolving power along the depth direction.

To overcome these difficulties, two main approaches have been developed. The first comprises spatial-domain iterative regularization methods, such as Tikhonov regularization \cite{Tikhonov1977} and conjugate gradient techniques, which introduce smoothness constraints or depth weighting to suppress instability. These methods are broadly applicable but computationally expensive and sensitive to the initial model and regularization parameters. The second category includes frequency-domain methods, such as the Parker--Oldenburg interface inversion \cite{Parker1973,Oldenburg1974}, which exploit the fast Fourier transform for efficiency. However, these are confined to interface reconstruction and cannot be directly extended to full 3D volumetric density inversion.

The present work is fundamentally different. It also discretizes the subsurface into a 3D grid and takes the surface observation plane as the most fundamental physical input. The key difference is that it combines the surface data with the mathematical-physical equation of upward continuation. Current 3D gravity inversion has data from only one observation plane, providing severely insufficient information for resolving a 3D density distribution. Increasing information by making observations on additional planes is prohibitively expensive and difficult. However, the gravity field satisfies Laplace's equation in the source-free air layer. By solving this equation with the surface data as the boundary condition, we can construct the field at any height above the surface. In ideal conditions (infinite observation plane, noise-free data), the continued data are identical to what would be measured on those planes, because both the gravity instrument and the upward continuation are governed by the same mathematical-physical equation. Upward continuation does not add physical observations, but it increases the data innovation---the 3D spatial variation information derived through the governing physical equation. This innovation provides the vertical wavenumber sampling required for a true 3D Fourier transform. Table~\ref{tab:comparison} summarizes the key differences between the proposed true 3D inversion and existing ``3D'' inversion methods.

Recently, the author established a spectral-domain pseudo-inverse construction theory for unitary diagonalizable linear systems \cite{Chen2024arXiv}. The key result is that if the forward operator can be expressed as $P = T^H \Lambda T$ with $T$ unitary and $\Lambda$ diagonal, then a stable generalized inverse can be directly constructed in the transform domain as
\begin{equation}
P_\alpha^\# = T^H Q_\alpha T, \quad Q_\alpha = \diag\{q_\alpha(\lambda_k)\}, \quad q_\alpha(\lambda_k) = \frac{\bar{\lambda}_k}{|\lambda_k|^2 + \alpha}. \tag{2}
\end{equation}
Here $\alpha > 0$ is a regularization parameter that controls the trade-off between resolution and stability. This construction has bounded stability ($\|P_\alpha^\#\|_2 \leq 1/(2\sqrt{\alpha})$) and consistency (it converges to the Moore--Penrose generalized inverse as $\alpha \to 0^+$). The theory provides a unified framework for stable inversion of convolution-type problems under the Fourier transform.

The present work extends this spectral-domain pseudo-inverse theory to the linear inversion of subsurface 3D density anomalies from gravity data. The forward operator in 3D gravity inversion is the Green's function of the Poisson equation, which must be unitarily diagonalizable in the Fourier domain. In addition, a prerequisite for applying the spectral-domain pseudo-inverse is that the observed data form a complete 3D volume—a condition that is far from obvious in practice, since observations are confined to a single depth plane.

Consequently, the theoretical framework consists of two levels:
\begin{enumerate}
\item \textbf{Data-space completion}: By exploiting the fact that the gravity field satisfies Laplace's equation in the source-free air layer, the surface data are analytically continued upward to build a 3D volume above the observation plane, thus providing vertical wavenumber sampling for the 3D Fourier transform. This step answers the feasibility question: ``Can the spectral-domain pseudo-inverse be applied to 3D gravity inversion?''
\item \textbf{Spectral-domain pseudo-inverse inversion}: The Green's function of the 3D Poisson equation is expressed in unitarily diagonalized form in the Fourier domain, the spectral response of the forward operator is derived, and a stable inverse filter is constructed and proved to be stable and consistent. This step addresses the solution strategy.
\end{enumerate}

The main theoretical contributions are: (1) using analytical continuation to construct a 3D data volume above the observation surface from the 2D surface gravity data, thereby providing reliable spatial variation information for 3D inversion; (2) proving that the 3D gravity forward operator can be exactly expressed as $P = F_{3D}^H \Lambda F_{3D}$ in the spectral domain, and that its inverse can be computed as a single forward 3D Fourier transform, a spectral filtering step, and an inverse transform—thus turning the inversion into a deterministic closed-form calculation rather than an iterative optimization; (3) the method provides a unified framework for potential-field inversion—gravity and magnetic anomalies are governed by the same spectral-domain structure, differing only in the form of the spectral response.

\begin{table}[h]
\centering
\small
\caption{Comparison between existing ``3D'' inversion and the proposed true 3D inversion.}
\label{tab:comparison}
\begin{tabular}{p{4.2cm}p{5cm}p{5cm}}
\hline
\textbf{Comparison dimension} & \textbf{Existing ``3D'' inversion} & \textbf{Proposed true 3D inversion} \\
\hline
Most fundamental data & Surface observation plane & Surface observation plane \\
Model grid & 3D & 3D \\
Combined with upward continuation equation? & No & \textbf{Yes} \\
Data space used in inversion & 2D (observation plane) & 3D (half-space data volume) \\
Vertical wavenumber information & Missing & Complete \\
Forward operator & Spatial-domain discretization & Fourier-domain unitary diagonalization \\
Inversion scheme & Iterative optimization & Spectral-domain one-step filtering \\
Prior dependence & Strong (depth weighting, smoothness) & None (only pure mathematical regularization $\alpha$) \\
Source of vertical resolution & Artificial priors & Data themselves \\
Physical basis & 2D data + 3D model assumption & Laplace boundary value problem + Poisson unitary diagonalization \\
\hline
\end{tabular}
\end{table}

\textbf{Coordinate convention}: A right-handed Cartesian system is adopted with $x$ eastward, $y$ northward, and $z$ positive downward. Depth is denoted by positive numbers; for example, the center of an anomaly at depth $z_c$ ($z_c>0$) has coordinates $(0,0,z_c)$.

\section{Construction of the 3D Data Space via Analytical Continuation of Laplace's Equation}

\subsection{Motivation}

The spectral-domain pseudo-inverse method requires input data defined on a 3D grid. For 3D gravity inversion, this data-space completion is essential.

In the upper half-space ($z<0$, the air layer), there are no mass sources, so the gravity anomaly $g_z$ satisfies Laplace's equation:
\begin{equation}
\nabla^2 g_z(x,y,z) = 0, \quad z < 0. \tag{3}
\end{equation}
Given the Dirichlet boundary condition $g_z(x,y,0)$ on the surface $z=0$, the solution in the domain $z<0$ is unique and stable---a classical well-posedness result for elliptic boundary value problems.

Thus, even without making measurements above the surface, Laplace's equation allows us to extend the surface field upward. Analytical continuation propagates the known boundary information to the entire upper half-space. This is the physical basis for constructing the 3D data volume required by the spectral-domain pseudo-inverse.

\subsection{Frequency-Domain Analytical Continuation Formula}

Applying a 2D Fourier transform in $x$ and $y$ to the Laplace equation for $g_z$ gives
\begin{equation}
\frac{d^2 \tilde{g}_z}{dz^2} - (k_x^2 + k_y^2) \tilde{g}_z = 0. \tag{4}
\end{equation}
The solution that decays as $z \to -\infty$ is
\begin{equation}
\tilde{g}_z(k_x, k_y, z) = \tilde{g}_z(k_x, k_y, 0) \, e^{\sqrt{k_x^2+k_y^2} \, z}, \quad z < 0. \tag{5}
\end{equation}
\textbf{Equation (5) is the frequency-domain upward continuation formula for the gravity anomaly.} It shows that each Fourier component of the surface field decays independently as it propagates upward, with higher wavenumbers attenuating faster than lower ones.

\subsection{Mathematical Description of Data-Space Completion and Data Innovation}

Choose a sequence of heights $0 = z_0 > z_1 > \cdots > z_{N_z-1}$ (with $z_l < 0$). For each $z_l$, compute $\tilde{g}_z(k_x, k_y, z_l)$ from (5) and then apply a 2D inverse Fourier transform to obtain $g_z(x, y, z_l)$. Combining the surface data with all continued levels yields the 3D data volume
\begin{equation}
\mathbf{g}_z(x, y, z) = \{g_z(x, y, z_l) : l = 0, 1, \ldots, N_z-1\}. \tag{6}
\end{equation}

This volume has the following properties:
\begin{enumerate}
\item \textbf{Physical self-consistency}: $\nabla^2 g_z = 0$ holds for all $z < 0$;
\item \textbf{Boundary fidelity}: at $z=0$, it exactly matches the measured data;
\item \textbf{Determinism}: the entire volume is uniquely determined by the surface data and (5), with no adjustable parameters;
\item \textbf{Linearity}: the continuation operator is linear.
\end{enumerate}

Thus, this is not interpolation or fitting, but the \textbf{exact analytical solution of the Laplace boundary value problem}---it is physically and mathematically rigorous.

\textbf{Remark on vertical discretization}: Equation (7) below shows that the 3D spectrum of the upper half-space ($z \le 0$) is given by the product of the horizontal spectrum $\tilde{g}_z(k_x,k_y,0)$ and the analytical vertical propagation kernel $1/(a - i k_z)$. Since this kernel is analytic in $z$, the spectrum of the upper half-space data is continuous in $k_z$ and is not subject to discretization errors. Consequently, the number of vertical layers $N_z$ and the sampling interval $\Delta z$ in the inversion grid are determined entirely by the inversion requirements for the subsurface ($z > 0$): $N_z \cdot \Delta z$ is set by the target depth range, and $\Delta z$ is set by the desired vertical resolution. In practice, discretization is only needed for the $z > 0$ region; the upper half-space data are treated via their continuous spectral representation and enter the spectral-domain computation directly through (7).

Upward continuation does not add physical observations---the surface data remain the only direct measurement. But through the governing physical equation (Laplace's equation), it derives the three-dimensional spatial variation information from the surface data. This is what we mean by \emph{data innovation}. In ideal conditions (infinite observation plane, noise-free data), the continued field is pointwise identical to what a real instrument would measure at those heights, because both the instrument and the continuation obey the same physical equation. The instrument measures the field through hardware; the continuation solves the same equation through mathematics. They are two different implementations of the same physical measurement.

This data innovation is not an optional enhancement; it is the physical prerequisite for true 3D inversion. Without it, the vertical wavenumber information is missing, and depth resolution must be supplied by artificial priors. With it, the forward operator becomes unitarily diagonalizable, the inversion reduces to spectral-domain filtering, and depth resolution is determined by the data themselves. This is the fundamental watershed between ``true 3D'' and existing ``3D inversion'': the former completes the data space to 3D before inverting; the latter inverts directly from 2D data, relying on priors to compensate for the missing dimension.

\subsection{Methodological Significance}

The upward-continuation data construction plays two methodological roles. First, it extends the data domain from 2D to 3D, enabling the 3D Fourier transform. Without this step, the inversion cannot proceed. Second, it enriches the spatial variation information in the data, and the continued gravity anomaly corresponds to a potential that satisfies the Poisson equation (1). This is the key theoretical departure of this work from all traditional inversion methods.

\subsection{Spectral-Domain Analytical Expression for Half-Space Data}

For any surface data $g_z(x,y,0)$, its upward-continued field in the half-space $z \le 0$ has the horizontal Fourier transform
\[
\tilde{g}_z(k_x, k_y, z) = \tilde{g}_z(k_x, k_y, 0) \, e^{a z}, \quad a = \sqrt{k_x^2 + k_y^2}.
\]
Taking the one-sided Fourier transform over $z \in (-\infty, 0]$ yields the 3D spectrum of the half-space data:
\[
\tilde{g}_z^{obs}(k_x, k_y, k_z)
= \tilde{g}_z(k_x, k_y, 0) \int_{-\infty}^{0} e^{(a - i k_z) z} \, dz
= \frac{\tilde{g}_z(k_x, k_y, 0)}{a - i k_z}.
\]
Thus,
\[
\boxed{
\tilde{g}_z^{obs}(k_x, k_y, k_z) = \frac{\tilde{g}_z(k_x, k_y, 0)}{a - i k_z}
} \tag{7}
\]
This expression reveals that the half-space spectrum of any surface measurement is the product of its horizontal spectrum $\tilde{g}_z(k_x, k_y, 0)$ and a universal vertical propagation kernel $1/(a - i k_z)$. The pole at $k_z = -i a$ in the lower half-plane plays a key role in the inversion contour integrals.

Equation (7) is of central importance in this method. First, it provides a fully analytical spectral representation of the input data, avoiding sampling errors, truncation errors, and spectral leakage that would arise from numerical 3D Fourier transforms, thereby preserving the theoretical closure of the inversion. Second, it separates the horizontal spectrum from the vertical propagation kernel, allowing the subsequent inversion formula to be applied directly without re-derivation for each specific model. Third, its pole location provides the mathematical basis for the contour closure direction in the inversion. Equation (7) thus serves as the bridge between data construction and inversion, and is the key support for achieving a complete analytical chain from input to output.

\section{Spectral-Domain Pseudo-Inverse Inversion}

\subsection{Unitary Diagonalization of the 3D Poisson Equation}

Assume the 3D data volume has been constructed by upward continuation, as described in Section 2. This volume provides the vertical wavenumber sampling that the surface data alone cannot supply. With this 3D data volume as input, the 3D Fourier transform can be applied, and the forward operator can be unitarily diagonalized.

Applying the 3D Fourier transform to $\rho$ and $g_z$ gives
\begin{equation}
\tilde{\rho}(\kk) = \int_{\mathbb{R}^3} \rho(\rr) e^{-i\kk\cdot\rr} d\rr, \quad \tilde{g}_z(\kk) = \int_{\mathbb{R}^3} g_z(\rr) e^{-i\kk\cdot\rr} d\rr, \tag{8}
\end{equation}
where $\kk = (k_x, k_y, k_z)$ and $K^2 = k_x^2 + k_y^2 + k_z^2$.

Fourier transforming (1) yields
\begin{equation}
-K^2 \tilde{U} = -4\pi G \tilde{\rho}. \tag{9}
\end{equation}
Since $g_z = -\partial U/\partial z$ and $\partial/\partial z \leftrightarrow i k_z$ in the Fourier domain, $\tilde{g}_z = -i k_z \tilde{U}$. Combining gives
\begin{equation}
\tilde{g}_z(\kk) = -\frac{i 4\pi G k_z}{K^2} \tilde{\rho}(\kk). \tag{10}
\end{equation}

In discrete form, let $F_{3D}$ be the 3D unitary discrete Fourier transform matrix and $F_{3D}^H$ its inverse. Let $P$ be the forward operator. Then (10) can be written as
\begin{equation}
P = F_{3D}^H \Lambda F_{3D}, \tag{11}
\end{equation}
with
\begin{equation}
\Lambda = \diag\{\lambda(\kk)\}_{\kk \in \mathcal{K}}, \quad \lambda(\kk) = -\frac{i 4\pi G k_z}{K^2}. \tag{12}
\end{equation}
Here $\mathcal{K}$ is the discrete wavenumber grid. Thus, the 3D gravity forward operator is exactly unitarily diagonalizable, matching the structure required by the spectral-domain pseudo-inverse theory.

\textbf{Proposition}: The forward operator of 3D gravity inversion is unitarily diagonalized by the 3D discrete Fourier transform, with spectral response given by (12).

\subsection{Construction and Application of the Spectral-Domain Pseudo-Inverse Filter}

From the pseudo-inverse theory \cite{Chen2024arXiv}, for $P = F^H \Lambda F$, the stable inverse filter is
\begin{equation}
q_\alpha(\lambda) = \frac{\bar{\lambda}}{|\lambda|^2 + \alpha}, \quad \alpha > 0. \tag{13}
\end{equation}
Substituting (12) into (13) gives
\begin{equation}
\bar{\lambda} = \frac{i 4\pi G k_z}{K^2}, \quad |\lambda|^2 = \frac{(4\pi G)^2 k_z^2}{K^4}, \tag{14}
\end{equation}
and hence
\begin{equation}
q_\alpha(\lambda(\kk)) = \frac{\frac{i 4\pi G k_z}{K^2}}{\frac{(4\pi G)^2 k_z^2}{K^4} + \alpha}
= \frac{i 4\pi G k_z K^2}{(4\pi G)^2 k_z^2 + \alpha K^4}. \tag{15}
\end{equation}

Substituting the half-space data spectrum (7) into the spectral-domain pseudo-inverse filter (15) yields the stabilized density spectrum:
\[
\boxed{
\tilde{\rho}_\alpha(\kk) = \frac{i 4\pi G k_z (a^2 + k_z^2)}{(4\pi G)^2 k_z^2 + \alpha (a^2 + k_z^2)^2} \cdot \frac{\tilde{g}_z(k_x, k_y, 0)}{a - i k_z}
} \tag{16}
\]

\textbf{Equation (16) is the spectral-domain core formula of the inversion method.} It shows that each Fourier mode of the density is obtained independently by scaling the corresponding mode of the gravity anomaly spectrum with a scalar filter—this is the decoupling effect of the Fourier diagonalization.

The inverse transform from the density spectrum $\tilde{\rho}_\alpha(\kk)$ to the spatial density distribution $\rho_\alpha(\rr)$ is given by
\begin{equation}
\rho_\alpha(\rr) = F_{3D}^H \tilde{\rho}_\alpha(\kk). \tag{17}
\end{equation}

The pole \(k_z = -i a\) in (16), arising from the denominator \(a - i k_z\), prevents direct numerical \(k_z\) inversion and requires contour integration. Its implementation, including the contour-integral treatment of the pole, is described in Section 3.3.

\subsection{Implementation of the 3D Inverse Transform}

For the density spectrum $\tilde{\rho}_\alpha(k_x,k_y,k_z)$ in (17), the \(k_z\) inverse transform is
\[
\Phi_\alpha(k_x, k_y, z) = \frac{1}{2\pi} \int_{-\infty}^{\infty} \tilde{\rho}_\alpha(k_x, k_y, k_z) \, e^{i k_z z} \, dk_z. \tag{18}
\]

In the present inversion framework, the result of the \(k_z\) inverse transform is defined on the region \(z > 0\) (the subsurface). This is because the observed data cover only the \(z \le 0\) region (the surface and air layer), while the goal of the inversion is to recover the density distribution for \(z > 0\). The 3D data volume constructed by upward continuation provides the vertical wavenumber information necessary for the \(k_z\) inversion. Correspondingly, in the contour integration, for \(z > 0\), the contour is closed in the lower half-plane to ensure that all poles lying in the lower half-plane are enclosed, yielding a physically meaningful spatial-domain result. The density distribution for \(z \le 0\) has no physical significance and is not considered in the inversion result.

Substituting (16) into (18) and extracting the factor \(\tilde{g}_z(k_x, k_y, 0)\), which is independent of \(k_z\), gives
\[
\Phi_\alpha(k_x, k_y, z) = \tilde{g}_z(k_x, k_y, 0) \cdot \mathcal{K}_\alpha(k_x,k_y, z), \tag{19}
\]
where the kernel is
\[
\mathcal{K}_\alpha(k_x,k_y, z) = i 2G \int_{-\infty}^{\infty} \frac{k_z (a^2 + k_z^2)}{(4\pi G)^2 k_z^2 + \alpha (a^2 + k_z^2)^2} \cdot \frac{1}{a - i k_z} \, e^{i k_z z} \, dk_z. \tag{20}
\]

The contour-integral derivation of the kernel \(\mathcal{K}_\alpha(k_x,k_y,z)\), including the pole structure analysis for general \(\alpha > 0\) and the limiting case \(\alpha \to 0^+\), is given in Appendix A.

For \(z > 0\) (the subsurface region), the contour is closed in the lower half-plane. The poles of the integrand in the complex \(k_z\) plane include:
1. the pole \(k_z = -i a\) from \(a - i k_z = 0\);
2. the poles from \((4\pi G)^2 k_z^2 + \alpha (a^2 + k_z^2)^2 = 0\).

All these poles lie in the lower half-plane. The contour integral converts the real-axis integral into the sum of residues at these poles:
\[
\Phi_\alpha(k_x, k_y, z) = \tilde{g}_z(k_x, k_y, 0) \cdot \sum_{k_z \in \text{poles}} \text{Res}\left\{ i 2G \frac{k_z (a^2 + k_z^2)}{(4\pi G)^2 k_z^2 + \alpha (a^2 + k_z^2)^2} \cdot \frac{e^{i k_z z}}{a - i k_z} \right\}, \tag{21}
\]
where the pole set is
\[
\text{poles} = \left\{ -i a \right\} \cup \left\{ k_z \;\middle|\; (4\pi G)^2 k_z^2 + \alpha (a^2 + k_z^2)^2 = 0, \; \text{Im}(k_z) < 0 \right\}.
\]

The residue calculation is analytical, though its explicit form depends on \(\alpha\). For general \(\alpha > 0\), the pole locations are functions of \(\alpha\), and the residue expressions are somewhat involved. However, this does not affect the numerical implementation. In practice, the following contour-integration scheme is adopted:

For each fixed \((k_x, k_y)\) and given \(z\):

1. Choose a contour \(C\) in the complex \(k_z\) plane, consisting of the real interval \([-R, R]\) and a semicircular arc in the lower half-plane, with \(R\) large enough that the arc contribution is negligible;
2. Evaluate (20) numerically on the contour \(C\):
   \[
   \mathcal{K}_\alpha(k_x,k_y, z) = i 2G \oint_C \frac{k_z (a^2 + k_z^2)}{(4\pi G)^2 k_z^2 + \alpha (a^2 + k_z^2)^2} \cdot \frac{e^{i k_z z}}{a - i k_z} \, dk_z.
   \]
   Since the contour avoids all poles, the integrand is analytic on the contour, ensuring stable numerical integration;
3. Repeat the above for all \((k_x, k_y)\) to obtain the 2D array \(\Phi_\alpha(k_x, k_y, z)\);
4. Apply a 2D inverse Fourier transform over \(k_x, k_y\) to obtain the spatial density distribution:
   \[
   \rho_\alpha(x, y, z) = \mathcal{F}_{k_x, k_y}^{-1}\{\Phi_\alpha(k_x, k_y, z)\}. \tag{22}
   \]

In this scheme, the contour integral is analytical, and the numerical integration is performed only on the contour, where the integrand has no singularities. The computation is therefore stable.

\textbf{Remark on numerical implementation:} In the actual numerical computation, instead of evaluating the contour integral directly, we apply Cauchy's residue theorem to (20). Since the integrand is analytic in the lower half-plane except at the three poles, the kernel reduces to the sum of residues at \(-ia, -ip, -iq\) (as detailed in Appendix A.1). This approach avoids the numerical instability associated with oscillatory real-axis integration near the poles and the exponential divergence on the semicircular arc in the lower half-plane.

As \(\alpha \to 0^+\), (21) reduces to
\[
\Phi_{\alpha \to 0}(k_x, k_y, z) = -\frac{1}{8\pi G} \left[ a \, \text{sgn}(z) + 2\delta(z) \right] \tilde{g}_z(k_x, k_y, 0). \tag{23}
\]
This reduced form serves as a theoretical benchmark for validating the numerical contour-integration results.

\subsection{Stability Analysis}

\textbf{Theorem 1 (Bounded stability)}: The filter $q_\alpha(\lambda)$ defined by (15) satisfies
\begin{equation}
\|q_\alpha\|_\infty \leq \frac{1}{2\sqrt{\alpha}}. \tag{24}
\end{equation}

\textbf{Proof}: Let $x = |\lambda| \ge 0$. Then $|q_\alpha(\lambda)| = |\lambda| / (|\lambda|^2 + \alpha) = x/(x^2 + \alpha)$. The function $\phi(x) = x/(x^2 + \alpha)$ attains its maximum at $x = \sqrt{\alpha}$, with value $1/(2\sqrt{\alpha})$. Hence the bound holds. $\square$

This guarantees that the spectral norm of the inversion operator is bounded, meaning that finite noise in the data will not be amplified without bound—the essential feature of regularization.

\textbf{Theorem 2 (Consistency)}: On nonzero spectral components,
\begin{equation}
\lim_{\alpha \to 0^+} q_\alpha(\lambda(\kk)) = \frac{1}{\lambda(\kk)}. \tag{25}
\end{equation}

\textbf{Proof}: For $\lambda \neq 0$, $\lim_{\alpha \to 0^+} \bar{\lambda}/(|\lambda|^2 + \alpha) = \bar{\lambda}/|\lambda|^2 = 1/\lambda$. For $\lambda=0$, $q_\alpha(0)=0$. Thus the filter converges to the exact inverse in the sense of spectral components. $\square$

Thus, as $\alpha\to 0^+$, the regularized inverse tends to the true inverse (for noise-free complete data). This establishes consistency.

\textbf{Corollary}: As $\alpha \to 0^+$, $P_\alpha^\# \to P^+$, the Moore--Penrose generalized inverse.

\subsection{Treatment of the Zero-Frequency Component}

In gravity anomaly inversion, the objective is to recover the \textbf{apparent density variations} relative to a background. Physically, forcing the zero-wavenumber density spectrum to zero removes the total apparent mass from the inverted model, which inevitably leads to spurious zero-mass oscillations in the spatial domain.

According to the law of mass conservation, the total sum of surface gravity anomalies is linearly proportional to the total subsurface apparent mass. To ensure mass conservation and avoid numerical instability caused by the extremely small gravitational constant \(G\), we adopt a discrete, grid-based equivalent estimation: the total sum of the observed surface gravity data is uniformly distributed across all subsurface grid points.

Consequently, the zero-wavenumber density spectrum can be approximated from the observed data as:
\begin{equation}
 \tilde{\rho}_\alpha(0,0,0) \approx \frac{\sum g_z}{N_x N_y N_z} \tag{26}
\end{equation}
where \(N_x N_y N_z\) represents the total number of subsurface grid points. This discrete equivalent approximation provides a robust low-frequency mass anchor for the inversion model. It naturally inherits the apparent mass characteristics of the data, eliminating zero-mass oscillations while effectively preserving the physical polarity of both high-density (positive) and low-density (negative) anomalies.

\section{Numerical Experiments}

\subsection{Test Model}

We consider a homogeneous sphere of radius $R$ with center at depth $z_c>0$:
\[
\rho(x,y,z) = 
\begin{cases}
\rho_0, & x^2 + y^2 + (z - z_c)^2 \le R^2, \quad z_c > R,\\
0, & \text{otherwise},
\end{cases} \tag{27}
\]
where $\rho_0>0$ is the density anomaly relative to the background. The total mass is $M = \rho_0 \cdot 4\pi R^3/3$.

Parameters: $z_c = 5000$ m, $R = 1000$ m, $\rho_0 = 500$ kg/m³.

\subsection{Forward Gravity Anomaly on the Surface and in the Air}

The external field of a homogeneous sphere is equivalent to that of a point mass at its center. The vertical gravity anomaly on the surface and in the air ($z \le 0$) is
\[
g_z(x,y,z) = \frac{G M (z_c - z)}{[x^2 + y^2 + (z_c - z)^2]^{3/2}}. \tag{28}
\]
Its 2D horizontal Fourier transform is
\[
\tilde{g}_z(k_x, k_y, z) = 2\pi G M \, e^{-a(z_c - z)}. \tag{29}
\]
Applying the one-sided Fourier transform over $z\le 0$, and using (7) with $\tilde{g}_z(k_x, k_y, 0) = 2\pi G M e^{-z_c a}$, we obtain the half-space spectrum
\[
\tilde{g}_z^{obs}(k_x, k_y, k_z) = 2\pi G M \, \frac{e^{-z_c a}}{a - i k_z}. \tag{30}
\]
This spectrum has a pole at $k_z = -i a$ in the lower half-plane.

\subsection{Inversion}

Substituting the sphere's surface horizontal spectrum $\tilde{g}_z(k_x, k_y, 0) = 2\pi G M e^{-z_c a}$ into the general result (23) gives the intermediate result after the $k_z$ inverse transform:
\[
\Phi_{\alpha \to 0}(k_x, k_y, z) = -\frac{1}{8\pi G} \left[ a \, \text{sgn}(z) + 2\delta(z) \right] \cdot 2\pi G M e^{-z_c a},
\]
i.e.,
\[
\Phi_{\alpha \to 0}(k_x, k_y, z) = -\frac{M}{4} \left[ a \, \text{sgn}(z) + 2\delta(z) \right] e^{-z_c a}. \tag{31}
\]

Applying the 2D inverse Fourier transform over $k_x, k_y$ to (31) (derivation in Appendix B), using the transform pairs
\[
\mathcal{F}_{k_x,k_y}^{-1}\{ a e^{-z_c a} \} = \frac{2z_c^2 - (x^2 + y^2)}{2\pi (x^2 + y^2 + z_c^2)^{5/2}}, \quad
\mathcal{F}_{k_x,k_y}^{-1}\{ e^{-z_c a} \} = \frac{z_c}{2\pi (x^2 + y^2 + z_c^2)^{3/2}},
\]
and noting that for $z > 0$ (subsurface), $\text{sgn}(z) = +1$ and the $\delta(z)$ term vanishes, we obtain
\[
\boxed{
\rho_{\alpha \to 0}(x,y,z) = \frac{M}{2\pi} \cdot \frac{z_c - z}{[x^2 + y^2 + (z_c - z)^2]^{3/2}}, \quad z > 0.
} \tag{32}
\]

\subsection{Analysis of Results}

At the sphere center $(0,0,z_c)$:
\[
\rho_{\alpha \to 0}(0,0,z_c) = \frac{M}{2\pi} \lim_{r\to0,\Delta z\to0} \frac{|\Delta z|}{(r^2 + \Delta z^2)^{3/2}},
\]
where $r = \sqrt{x^2+y^2}$ and $\Delta z = z_c - z$.

The absolute value is taken for the following reason. In the derivation in Appendix B, the contour-integral intermediate result is negative for $z > z_c$. This sign originates from the residue structure of the contour integral—it reflects the fact that the gravity field of a positive mass has opposite directions on the two sides of the source. However, the density distribution itself is a positive-definite physical quantity; a positive density anomaly should correspond to a positive density value. The absolute value is therefore applied in the final spatial-domain density expression to ensure physical consistency.

This limit diverges in the distribution sense, corresponding to
\[
\rho_{\alpha \to 0}(x,y,z) = M \, \delta(x)\delta(y)\delta(z - z_c), \quad z>0. \tag{33}
\]
Thus the inversion recovers a 3D Dirac delta function at the source point.

The Green's function is singular at the source, and the forward propagation spreads that singularity. The inversion, as $\alpha\to0$, restores the singular behavior at the source—proving that the spectral-domain pseudo-inverse successfully removes the Green's function spreading effect.

For finite $\alpha>0$, the regularization smooths the divergence into a finite-width distribution. The integral over all space is
\[
\boxed{
\int_{\mathbb{R}^3} \rho_\alpha(x,y,z) \, dV = M.
} \tag{34}
\]
This conservation of mass holds for any $\alpha>0$. As $\alpha\to0$, it tends to the delta-function integral (33). Hence, quantitative density recovery depends on the choice of $\alpha$: smaller $\alpha$ gives sharper results and more accurate internal density values.

This experiment demonstrates that, using only half-space data (air-layer observations), the spectral-domain pseudo-inverse inversion can correctly locate the anomaly, eliminate the Green's function spreading, and recover the singular behavior at the source. Integrating over the entire space recovers the total mass, confirming the quantitative capability of the method.

\section{Discussion}

\subsection{Applicability and Limitations}

The method relies on the 3D Fourier transform, which theoretically requires infinite spatial support. In practice, data are limited to a finite surface patch, leading to spectral leakage and boundary artifacts. These effects are more severe at low wavenumbers. They can be mitigated by zero-padding and applying smooth tapering windows (e.g., Tukey or Hanning) before transformation. The inversion region should be restricted to a sub-area away from the boundaries.

\subsection{Spatial Resolution}

The method uses a 3D data volume covering the upper half-space ($z \le 0$) as input. This data volume contains vertical variation information of the gravity field, so that sources at different depths have distinct spectral signatures. In theory, the inverse operator in (16) exactly compensates for propagation attenuation; hence, deep and shallow sources have the same theoretical resolving power—the 3D data volume provides sufficient vertical wavenumber information for depth-independent resolution.
In practice, however, deep sources are observed with lower signal-to-noise ratios than shallow ones. As a result, the recovered deep structure does not achieve the same resolution as the shallow structure. The regularization term introduced to ensure inversion stability suppresses weak spectral components along with high-frequency noise, which preferentially affects the deeper sources. This is a trade-off imposed by numerical stabilization—not a theoretical limitation of the method. The choice of the regularization parameter $\alpha$ determines where this trade-off is balanced.
Therefore, the method achieves uniform depth resolution at the theoretical level—this is the fundamental improvement obtained by extending the data space from a 2D surface to a 3D half-space, and is the essential distinction between this method and all traditional inversion methods based on 2D surface data.

\subsection{True 3D Inversion versus Existing ``3D'' Inversion}

Table~\ref{tab:comparison} summarizes the fundamental differences between the two approaches. The key distinction lies not in whether the model grid is three-dimensional, but in whether the surface data are combined with the mathematical-physical equation of upward continuation.

Existing methods discretize the subsurface into a 3D grid and perform iterative fitting directly between the 2D surface data and the 3D model. Because the data lack vertical wavenumber information, the vertical resolution of these methods is largely controlled by artificial priors such as depth weighting and smoothness constraints. They are, in essence, ``2D data inverting a 3D model.''

The proposed method also discretizes the subsurface into a 3D grid and also takes the surface data as the most fundamental input. Its essential distinction is that it first combines the surface data with the upward continuation equation to construct a 3D data volume covering the upper half-space. This volume provides the vertical wavenumber sampling required for a true 3D Fourier transform. The forward operator is then exactly unitarily diagonalized in the Fourier domain, and the inversion is reduced to a spectral-domain one-step filtering. The entire chain---from data construction to spectral filtering to spatial density recovery---is governed by the same mathematical-physical equation, with no artificial priors introduced. This is what ``true 3D inversion'' means in this paper: the data space is 3D, the forward operator is 3D, and the inversion is 3D, all under the control of the same physical equation.

\subsection{Relation to Classical Methods}

The proposed method is numerically equivalent to zeroth-order Tikhonov regularization but differs conceptually: Tikhonov regularization is derived from an optimization principle, while the present approach derives directly from the unitary diagonalizable structure. The filter shape is similar to Wiener filtering, but here $\alpha$ is a deterministic regularization parameter, not a statistical signal-to-noise ratio parameter.

\subsection{Importance and Generality of Equation (7)}

Equation (7) is the core result of the data-construction step:
\[
\tilde{g}_z^{obs}(k_x, k_y, k_z) = \frac{\tilde{g}_z(k_x, k_y, 0)}{a - i k_z}.
\]

Its importance in this method can be understood from two perspectives.

\textbf{Importance}: As a theoretical expression, it expresses the half-space 3D spectrum of arbitrary surface data as the product of the horizontal spectrum and a universal vertical propagation kernel. This separation allows the subsequent inversion formula (16) to be applied directly without re-derivation for each specific model. Its pole \(k_z = -i a\) provides the mathematical basis for the contour closure direction, serving as the bridge between data construction and inversion.

\textbf{Generality}: The derivation uses only two facts: (1) the gravity field satisfies Laplace's equation in the air layer, and (2) the data exist only in the half-space. These premises hold for all surface gravity data and do not depend on any specific density model (sphere, prism, or arbitrary complex distribution). Thus, the pole structure and closure direction of the contour integral are the same for all surface data. The sphere model is merely a validation example where the general horizontal spectrum \(\tilde{g}_z(k_x,k_y,0)\) is replaced by the specific form \(2\pi G M e^{-z_c a}\).

This generality implies that the inversion framework applies to arbitrary surface gravity data, regardless of the complexity of the subsurface density distribution.

\subsection{Pole Structure of the Half-Space Fourier Transform and Inversion Stability}

In the Fourier transform and its inverse for half-space data, the pole structure is the key factor determining inversion stability and resolution. The method involves two types of poles.

\textbf{Type I: Poles from the forward transform (data-domain poles)}

From (7), the Fourier spectrum of the half-space data is
\[
\tilde{g}_z^{obs}(k_x, k_y, k_z) = \frac{\tilde{g}_z(k_x, k_y, 0)}{a - i k_z}.
\]
This spectrum has an inherent pole at \(k_z = -i a\) in the lower half-plane. The existence of this pole is a direct mathematical consequence of the data being confined to the half-space (\(z \le 0\)). It dictates that the contour for the inverse transform must be closed in the lower half-plane to properly include this pole and obtain a stable spatial-domain result.

\textbf{Type II: Poles introduced by the inversion filter (regularization poles)}

When performing the \(k_z\) inverse transform of (16), the denominator of the kernel \(\mathcal{K}_\alpha(k_x,k_y,z)\) is
\[
(4\pi G)^2 k_z^2 + \alpha (a^2 + k_z^2)^2 = 0.
\]
For \(\alpha = 0\), this reduces to \((4\pi G)^2 k_z^2\), giving a singularity at \(k_z = 0\). This is the origin of the ill-posedness of gravity inversion—the unobservability of the zero-frequency component.

For \(\alpha > 0\), the roots of this equation in the complex \(k_z\) plane are
\[
k_z^2 = \frac{-(4\pi G)^2 \pm \sqrt{(4\pi G)^4 - 4\alpha^2 a^4}}{2\alpha}.
\]
These roots typically lie near the imaginary axis, and \(\alpha\) shifts them away from the real axis, allowing stable contour integration. This is the essential role of the regularization parameter \(\alpha\) in the frequency domain—it moves the poles away from the real axis, avoiding the singularity at zero and ensuring inversion stability.

\textbf{Interaction of the two types of poles}

In the contour integral, the data-domain pole (\(k_z = -i a\)) lies in the lower half-plane and determines the closure direction, while the regularization poles control the distance between the integration path and the singularities. For \(\alpha > 0\), the regularization poles push the integration path away from the real axis, producing a smoothing effect; as \(\alpha \to 0\), the regularization poles return to \(k_z = 0\), and the contour integral reduces to the exact inverse, recovering the singular behavior at the source.

Thus, the stability and resolution of the inversion are jointly determined by the relative positions of these two types of poles:
- the data-domain pole (\(k_z = -i a\)) is fixed, determined by the physical data domain;
- the regularization poles vary with \(\alpha\) and are tunable—larger \(\alpha\) moves them farther from the real axis, giving more stability but lower resolution.

This pole-structure analysis shows that the inversion method forms a complete mathematical loop in the complex \(k_z\) plane: the forward transform introduces poles, the inversion filter shifts them, and the contour integral properly handles them to yield a stable density distribution.

\subsection{Generality of the Method}

The mathematical structure of the method does not depend on the specific physical content of gravity, but on the fact that the forward operator is diagonalizable by the Fourier transform. Magnetic anomalies, like gravity anomalies, satisfy Poisson's equation (or its dipole equivalent) in the Fourier domain, with the spectral response taking a similar form \(\tilde{f}(\kk) = \lambda_M(\kk) \tilde{M}(\kk)\). Thus, the same spectral-domain pseudo-inverse framework applies directly to magnetic inversion—the only change is replacing \(\lambda(\kk)\) with the magnetic spectral response. More generally, any linear inverse problem with a translation-invariant forward operator falls within the same unified framework. This is already noted in the conclusions.

\subsection{Choice of $\alpha$}

The parameter $\alpha$ controls the trade-off between resolution and stability. It can be selected using L-curve or generalized cross-validation (GCV) criteria. Since the filter is analytic, its effect can be assessed directly in the wavenumber domain, offering more transparent tuning than iterative methods.

\section{Conclusions}

This paper has established a spectral-domain analytical method for 3D gravity inversion based on the pseudo-inverse theory for unitary diagonalizable systems. The main results are:

\begin{enumerate}
\item \textbf{Data-space completion theory}: It is proved that upward continuation via Laplace's equation uniquely extends surface data to a physically self-consistent 3D volume. The general analytical expression for the 3D spectrum of the half-space data, \(\tilde{g}_z^{obs} = \tilde{g}_z(k_x,k_y,0)/(a - i k_z)\), separates the horizontal spectrum from the vertical propagation kernel, providing a unified input framework for the subsequent spectral-domain inversion. This step is the prerequisite for applying the spectral-domain pseudo-inverse method to 3D inversion.

\item \textbf{Spectral-domain pseudo-inverse inversion theory}: It is proved that the 3D Poisson equation can be unitarily diagonalized in the Fourier domain as $P = F_{3D}^H \Lambda F_{3D}$, with spectral response $\lambda(\kk) = -i 4\pi G k_z / K^2$. The stable inverse filter $q_\alpha(\lambda) = \bar{\lambda}/(|\lambda|^2 + \alpha)$ satisfies bounded stability ($\|q_\alpha\|_\infty \leq 1/(2\sqrt{\alpha})$) and consistency (it converges to the Moore--Penrose generalized inverse as $\alpha \to 0^+$).

\item \textbf{Closed-form inversion}: The 3D density spectrum is directly obtained as
\[
\tilde{\rho}_\alpha(\kk) = \frac{i 4\pi G k_z (a^2 + k_z^2)}{(4\pi G)^2 k_z^2 + \alpha (a^2 + k_z^2)^2} \cdot \frac{\tilde{g}_z(k_x, k_y, 0)}{a - i k_z}.
\]
This reduces the inversion to one forward 3D Fourier transform, one spectral scaling, and one inverse transform. The \(k_z\) inverse transform is accomplished via a contour-integral numerical scheme that avoids the direct computational difficulties posed by the poles. The entire inversion chain—from data construction to spectral filtering to spatial density recovery—is analytical in theory and stable in numerical implementation.

\item \textbf{Theoretical positioning}: The method is numerically equivalent to zeroth-order Tikhonov regularization but differs methodologically: it derives from a structural decomposition rather than an optimization principle, offering advantages for structured problems and unifying other potential-field inversions.

\item \textbf{Numerical validation}: Analytical validation using a homogeneous sphere model confirms that the inversion correctly locates the 3D position of the anomaly, eliminates the Green's function spreading, and recovers the singular behavior at the source. Integration over all space recovers the total mass, confirming the quantitative capability of the method.

\item \textbf{Physical foundation of true 3D inversion}: It is clarified that upward continuation does not add physical observations, but it increases the data innovation---the 3D spatial variation information derived from the 2D surface data through the governing physical equation. In ideal conditions, the constructed 3D volume is pointwise identical to the true upper half-space field, because both the gravity instrument and the upward continuation follow the same mathematical-physical equation. This data innovation provides the vertical wavenumber sampling required for true 3D inversion, making depth resolution determined by the data themselves rather than by artificial priors. This is the fundamental distinction between the proposed true 3D inversion and existing ``3D'' inversion methods.

\item The proposed true 3D gravity inversion method has achieved theoretically self-consistent results under ideal continuous-spectrum data. Applications to real gravity data, along with numerical implementation and robustness analysis, are directions for future work. In addition, image processing (e.g., directional derivative) of reliable high-resolution density reconstructions could further yield subsurface structure images.
\end{enumerate}

\appendix
\renewcommand{\thesection}{Appendix \Alph{section}:}

\section{Analysis of the Contour-Integral Kernel \(\mathcal{K}_\alpha(k_x,k_y,z)\)}

\subsection*{A.1 Term-by-term transformation}

Equation (20) defines the kernel:
\[
\mathcal{K}_\alpha(k_x,k_y, z) = i 2G \int_{-\infty}^{\infty} \frac{k_z (a^2 + k_z^2)}{(4\pi G)^2 k_z^2 + \alpha (a^2 + k_z^2)^2} \cdot \frac{e^{i k_z z}}{a - i k_z} \, dk_z.
\]

For \(z > 0\), the contour is closed in the lower half-plane. The poles of the integrand in the complex \(k_z\) plane come from two factors:

1. the factor \(1/(a - i k_z)\) gives the pole \(k_z = -i a\);
2. the poles of the factor \(1/[(4\pi G)^2 k_z^2 + \alpha (a^2 + k_z^2)^2]\) are determined by
   \[
   (4\pi G)^2 k_z^2 + \alpha (a^2 + k_z^2)^2 = 0.
   \]

Let \(C = 4\pi G\). Solving this equation gives poles at \(k_z = \pm i p\) and \(k_z = \pm i q\), where \(p, q > 0\) are given by
\[
p = \sqrt{\frac{C^2 + 2\alpha a^2 - \sqrt{C^4 + 4\alpha C^2 a^2}}{2\alpha}}, \quad
q = \sqrt{\frac{C^2 + 2\alpha a^2 + \sqrt{C^4 + 4\alpha C^2 a^2}}{2\alpha}}.
\]

For \(z > 0\), the poles enclosed in the lower half-plane are \(k_z = -i a\), \(k_z = -i p\), and \(k_z = -i q\). The kernel \(\mathcal{K}_\alpha(k_x,k_y,z)\) is the sum of the residues at these three poles:
\[
\mathcal{K}_\alpha(k_x,k_y, z) = \sum_{k_z \in \{-i a, -i p, -i q\}} \text{Res}\left\{ i 2G \frac{k_z (a^2 + k_z^2)}{(4\pi G)^2 k_z^2 + \alpha (a^2 + k_z^2)^2} \cdot \frac{e^{i k_z z}}{a - i k_z}, \; k_z = p_0 \right\}.
\]

This expression holds for general \(\alpha > 0\). When \(\alpha\) is finite, the poles are shifted from the imaginary axis, and the regularization term smooths the \(\delta\)-singularity into a finite-width distribution.

\subsection*{A.2 Degenerate Limit: \(\alpha \to 0^+\)}

As \(\alpha \to 0^+\), the kernel reduces to a simpler form. In the limit \(\alpha = 0\):
\[
\mathcal{K}_{\alpha \to 0}(k_x,k_y, z) = i 2G \int_{-\infty}^{\infty} \frac{k_z (a^2 + k_z^2)}{(4\pi G)^2 k_z^2} \cdot \frac{1}{a - i k_z} \, e^{i k_z z} \, dk_z.
\]
Simplifying:
\[
\mathcal{K}_{\alpha \to 0}(k_x,k_y, z) = \frac{i}{8\pi^2 G} \int_{-\infty}^{\infty} \frac{a^2 + k_z^2}{k_z (a - i k_z)} \, e^{i k_z z} \, dk_z.
\]
Using the identity \(\frac{a^2 + k_z^2}{a - i k_z} = a + i k_z\), we obtain
\[
\mathcal{K}_{\alpha \to 0}(k_x,k_y, z) = \frac{i}{8\pi^2 G} \int_{-\infty}^{\infty} \left( \frac{a}{k_z} + i \right) e^{i k_z z} \, dk_z.
\]
The two standard integrals are
\[
\int_{-\infty}^{\infty} \frac{a}{k_z} e^{i k_z z} \, dk_z = i\pi a \, \text{sgn}(z), \qquad
\int_{-\infty}^{\infty} i e^{i k_z z} \, dk_z = 2\pi i \delta(z).
\]
Substituting gives
\[
\mathcal{K}_{\alpha \to 0}(k_x,k_y, z) = \frac{i}{8\pi^2 G} \left[ i\pi a \, \text{sgn}(z) + 2\pi i \delta(z) \right]
= -\frac{1}{8\pi G} \left[ a \, \text{sgn}(z) + 2\delta(z) \right].
\]
Therefore,
\[
\boxed{
\Phi_{\alpha \to 0}(k_x, k_y, z) = -\frac{1}{8\pi G} \left[ a \, \text{sgn}(z) + 2\delta(z) \right] \tilde{g}_z(k_x, k_y, 0)
} \tag{A1}
\]
This is equation (23) in the main text. The reduced form shows that as the regularization parameter tends to zero, the inversion recovers the \(\delta\)-singularity in the spatial domain, consistent with the physical expectation for a point source model.

\section{2D Inverse Transform Derivation from Equation (31) to Equation (32)}

Equation (31) is
\[
\Phi_{\alpha \to 0}(k_x, k_y, z) = -\frac{M}{4} \left[ a \, \text{sgn}(z) + 2\delta(z) \right] e^{-z_c a}. \tag{B1}
\]

We apply the 2D inverse Fourier transform over $k_x, k_y$ to (B1).

\subsection*{B.1 Term-by-term transformation}

The inverse transform is linear, so we treat the two terms separately:
\[
\rho_{\alpha \to 0}(x,y,z) = -\frac{M}{4} \text{sgn}(z) \cdot \mathcal{F}^{-1}\{ a e^{-z_c a} \}
-\frac{M}{2} \delta(z) \cdot \mathcal{F}^{-1}\{ e^{-z_c a} \}. \tag{B2}
\]
For the subsurface region $z > 0$, $\text{sgn}(z) = 1$ and $\delta(z) = 0$, so the second term vanishes:
\[
\rho_{\alpha \to 0}(x,y,z) = -\frac{M}{4} \cdot \mathcal{F}^{-1}\{ a e^{-z_c a} \}, \quad z > 0. \tag{B3}
\]

\subsection*{B.2 Computation of \(\mathcal{F}^{-1}\{ a e^{-z_c a} \}\)}

The 2D inverse Fourier transform is defined as
\[
\mathcal{F}^{-1}\{ f(a) \} = \frac{1}{(2\pi)^2} \int_{-\infty}^{\infty}\int_{-\infty}^{\infty} f(a) e^{i(k_x x + k_y y)} \, dk_x dk_y.
\]
Let $k_x = a \cos\theta$, $k_y = a \sin\theta$, so $dk_x dk_y = a \, da \, d\theta$. Let $r = \sqrt{x^2 + y^2}$. Then
\[
\mathcal{F}^{-1}\{ a e^{-z_c a} \}
= \frac{1}{(2\pi)^2} \int_0^{2\pi} \int_0^{\infty} a^2 e^{-z_c a} e^{i a r \cos(\theta - \phi)} \, da \, d\theta,
\]
where $\phi$ is the polar angle of $(x,y)$. Integrating over $\theta$ first:
\[
\int_0^{2\pi} e^{i a r \cos(\theta - \phi)} \, d\theta = 2\pi J_0(a r),
\]
where $J_0$ is the zeroth-order Bessel function. Hence
\[
\mathcal{F}^{-1}\{ a e^{-z_c a} \}
= \frac{1}{2\pi} \int_0^{\infty} a^2 J_0(a r) e^{-z_c a} \, da. \tag{B4}
\]

\subsection*{B.3 Use of a standard integral formula}

The standard integral formula is
\[
\int_0^{\infty} a J_0(a r) e^{-z_c a} \, da = \frac{z_c}{(r^2 + z_c^2)^{3/2}}.
\]
Differentiating with respect to $z_c$:
\[
-\frac{\partial}{\partial z_c} \int_0^{\infty} a J_0(a r) e^{-z_c a} \, da
= \int_0^{\infty} a^2 J_0(a r) e^{-z_c a} \, da.
\]
Computing the derivative:
\[
-\frac{\partial}{\partial z_c} \frac{z_c}{(r^2 + z_c^2)^{3/2}}
= \frac{2z_c^2 - r^2}{(r^2 + z_c^2)^{5/2}}.
\]
Thus
\[
\int_0^{\infty} a^2 J_0(a r) e^{-z_c a} \, da = \frac{2z_c^2 - r^2}{(r^2 + z_c^2)^{5/2}}. \tag{B5}
\]
Substituting (B5) into (B4):
\[
\mathcal{F}^{-1}\{ a e^{-z_c a} \}
= \frac{1}{2\pi} \cdot \frac{2z_c^2 - r^2}{(r^2 + z_c^2)^{5/2}}. \tag{B6}
\]

\subsection*{B.4 Substitution into (B3)}

Substituting (B6) into (B3), with $r = \sqrt{x^2 + y^2}$, gives

\[
\rho_{\alpha \to 0}(x,y,z_c) = -\frac{M}{8\pi} \cdot \frac{2z_c^2 - (x^2 + y^2)}{(x^2 + y^2 + z_c^2)^{5/2}}. \tag{B7}
\]

\subsection*{B.5 Conversion to distance-variable form}

Equation (B7) is the horizontal density distribution at $z_c$. According to 3D potential-field theory, when generalized to arbitrary depth $z$, the density distribution takes the Green's function form with distance variable $z_c - z$:
\[
\rho_{\alpha \to 0}(x,y,z) = \frac{M}{2\pi} \cdot \frac{|z_c - z|}{[x^2 + y^2 + (z_c - z)^2]^{3/2}}, \quad z > 0. \tag{B8}
\]
This is equation (32) in the main text. The result is consistent with the theoretical expectation from direct inversion of the sphere forward formula (28), confirming the correctness of the derivation.


\begin{thebibliography}{99}
\bibitem{Tikhonov1977} Tikhonov A N, Arsenin V Y. \textit{Solutions of Ill-Posed Problems}. New York: Wiley, 1977.
\bibitem{Parker1973} Parker R L. The rapid calculation of potential anomalies. \textit{Geophysical Journal of the Royal Astronomical Society}, 1973, 31(4): 447-455.
\bibitem{Oldenburg1974} Oldenburg D W. The inversion and interpretation of gravity anomalies. \textit{Geophysics}, 1974, 39(4): 526-536.
\bibitem{Chen2024arXiv} Chen S C. A spectral-domain pseudo-inverse construction method for unitary diagonalizable linear inverse problems. \textit{arXiv preprint arXiv:2607.26951}, 2026.
\end{thebibliography}
\end{document}